\documentclass[twocolumn,preprintnumbers,amsmath,amssymb,showpacs,pra]{revtex4-1}
\usepackage{graphicx}
\usepackage{mathrsfs}
\usepackage[colorlinks]{hyperref}
\usepackage{dcolumn}
\usepackage{bm}

\begin{document}
%\preprint{APS/123-QED}
\title{Distortion of Interference Fringes and the Resulting Vortex Production of Merging Bose-Einstein Condensates}
\author{Bo Xiong$^1$, Tao Yang$^{2,*}$ and Keith A. Benedict$^1$}
%\altaffiliation{Present Address: Seoul National University, Department of Physics and Astronomy, Center for Theoretical Physics, 151-747 Seoul, Korea}
\affiliation{$^1$Midlands Ultracold Atom Research Center, School of Physics and Astronomy, University of Nottingham, Nottingham NG7 2RD, United Kingdom\\
$^2$Institute of Modern Physics, Northwest University, Xi'an 710069, P. R. China}

\date{\today}
\email{yangt@nwu.edu.cn}
\begin{abstract}
We investigate the effects of interatomic interactions and expansion on the distortion of interference fringes of a pair of initially well-separated, but coherent, condensate clouds trapped in a harmonic trap. The distortion of interference fringes, which can lead to the spontaneous formation of vortices in the atom clouds, depends crucially on two relevant parameters: the center-of-mass velocity and peak density of the initial state. We identify three qualitatively distinct regimes for the interfering condensates: collision, expansion, and merging, by the spatial and temporal features of the fringe spacings. Using a comprehensive set of numerical simulations based on the Gross-Pitaevskii equation, we specify the cross-overs between these regimes and  propose the optimal the system parameters required for dynamical instabilities and vortex creation.
\end{abstract}

\pacs{03.75.Dg, 03.75.Lm, 67.85.-d}

\maketitle
\section{Introduction}
Interference of two spatially separated Bose condensed clouds of ultra-cold atoms has been widely studied \cite{Andr, stock, shin, shin1, davi, schu, bloc, hoff, hadz1, pete, Jo, Jo1, PRA.87.023603}. These researches are of fundamental interest, for example, in demonstrating the quantum nature of the condensate and in investigating decoherence. In addition, such processes center on matter wave interferometry using ultra-cold atom condensates, which may have many technological applications. Typically, there are two limiting cases in which initially separated clouds can be made to evolve so that they subsequently overlap. Firstly, they can be allowed to expand by releasing from the confining potential that holds them apart, i.e. free expansion \cite{Andr, shin1, shin}, to avoid the complicating effects of interactions. Secondly, they can be subjected to external potential gradients that cause the clouds to move together and collide whilst maintaining their form \cite{Jo, Jo1, PRA.87.023603}. In realistic scenarios, a interference process will, to some extent, involve both the shape distortion and center-of-mass (c.m.) oscillation by keeping but relaxing the trap potential. The resulting interference pattern will be more complex than in either of the idealized cases, particularly for systems in which inter-atomic interactions are involved. We refer these three cases as expansion, collision and merging, respectively. There has been some theoretical work on freely expanding condensate clouds in which inter-atomic interactions are shown to increase the expansion rate \cite{wall} and create nonuniform interference fringes \cite{ming, rohr}. However, to date there is little study on the general case, i.e. merging, or how it compares with either the purely expanding or purely colliding scenarios; in particular, the complex dynamics of curved fringes in the interfering BECs has not been systematically studied.

It is now known that, while the behavior of very low density condensates can be well described using elementary single-particle quantum mechanics, systems in which the interactions are stronger show behavior which is quantitatively, and frequently qualitatively, different. It was shown in Ref.\cite{Scott} that the interference of two counterpropagating interacting clouds can give rise to the formation of persistent dark solitary waves and, subsequently, the nucleation of linear arrays of vortex rings. As well as having implications for real matter wave interferometers, these processes are of intrinsic interest as an experimentally controllable route to homogeneous quantum turbulence \cite{vinen}. Recent experiment \cite{Sche} has observed the spontaneous formation of vortices in a ring trap in which the condensate is split into three components and then allowed to recombine. Related theoretical work focuses on the role of recombination time (controlled by the time of ramping down the central barrier of the trap potential) and relative phase difference (controlled by the holding time or phase imprinting\cite{PRA.60.3381}) on the formation of these spontaneous vortices \cite{Carr, Carr1, PRA.87.023603}. However, the detailed description of how the combination of expansion and interference of atom clouds affects the formation of nonlinear excitations has not been shown.

In this paper, we model experiments in which a condensate is divided adiabatically into two coherent clouds, for example by using a tailored magnetic potential from an atom chip \cite{shin1}, the introduction of an optical barrier via a shaped blue-detuned laser \cite{davi} or by passing counterpropagating red detuned laser beams through an acousto-optic modulator driven at RF frequencies \cite{shin}. We examine how the process of both expansion and collision affects the interference patterns by varying the initial peak densities of the clouds and their relative velocities of c.m. oscillations (controlled by varying the trapping potential in which they move). The situation of interference with respect to different initial displacements of condensates has been well studied in Ref.\cite{PRA.87.023603}, while the shape distortion of the clouds does not contribute. The system is driven to different regimes by the competition of the c.m. kinetic energy and the interatomic interactions. In the underdamped regime the c.m. kinetic energy is much larger than the interaction energy and the system responds linearly similar to the noninteracting system, while the overdamped regime is on the contrary and the system is dominated by nonlinear effects. The critically damped regime is a intermediate regime where those two energy are of the same order. In this paper we analyze the effects of combination of c.m. oscillation and shape distortion on the system dynamics and excitation properties. Our numerical simulations, based on the zero-temperature mean-field Gross-Pitaevskii equation (GPE), show that in the overdamped regime, interference fringes first develop non-uniform spacings and then become curved. In extreme cases, the curvature can become sufficient for there to be a net circulation around a localized core region leading to the formation of a vortex ring. This can be achieved at high densities but low c.m. velocities during interference. While the maximum c.m. velocities is high enough, the relative interference time is short and the system cannot respond hydrodynamically to the high-density fringes before the clouds have passed through one another and the fringes have disappeared. Through the analysis of a large set of simulations, we argue that there is a generic mechanism by which the curvature of interference fringes due to interatomic interactions and spatial non-uniformity can produce a net circulation and vortex formation. We further identify a locus of points in parameter space at which the instability fringe appears.

This paper is organized as follows. In section II we describe the models we used to construct the system of two interfering condensates in three different cases, i.e. expanding, colliding and merging. The numerical procedure for simulating the dynamics of the given system is also introduced in this section. In section III, we propose a phenomenological formula which quantitatively describes the interference pattern of two condensates. We also identify three distinct interfering cases of the system dynamics. Section IV shows the typical interference patterns of the three different processes and the formation of spontaneous vortices in merging condensates. Also the relation between vortex formation and interference in merging process is discussed. In section V, we summarize general properties of the parameter space, including the onset of the instability and vortex formation. Section VI contains our conclusions.

\section{Simulation models}
Our general protocol involves preparing each of a pair of clouds in the lowest energy state of a harmonic trap.
The prepared clouds are then displaced in opposite directions and allowed to evolve when subject to a harmonic trap potential, which need not be the same as that used to prepare the initial clouds. We begin with a trap potential of the form
    \begin{equation} \label{Eq.1}
      V(r,z;t) = \frac{1}{2}m\omega_{\perp}^{2}(t)r^2 + \frac{1}{2}m \omega_{z}^{2}(t)z^2\qquad,
    \end{equation}
where $r=\sqrt{x^2+y^2}$ and $z$ are space coordinates of positions within the trap. The initial wave function $\phi_{0}(r,z)$ can be obtained by minimizing the energy functional
    \begin{eqnarray} \label{Eq.2}
            E[\phi_{0}] &= 2\pi \int_{-\infty}^{\infty}dz \int_0^{\infty}rdr  \left\{ \frac{\hbar^2}{2m} \left\vert\nabla\phi_{0}\right\vert^2   + V({r},z;0)\left\vert\phi_{0}\right\vert^2\right.\nonumber\\
                       &~~~+\left.\frac{1}{2} g\left\vert\phi_{0}\right\vert^4 \right\} ,
    \end{eqnarray}
where $g=4\pi\hbar^2 a_s/m$ is the inter-atomic coupling constant, and $a_s$ is the $s$-wave scattering length. We use the values $a_s=2.9$nm and $m=3.82 \times 10^{-26}$kg appropriate to a Sodium-23 condensate throughout this paper. The corresponding normalization condition is
    \begin{equation} \label{Eq.3}
      N= 2\pi \int_{-\infty}^{\infty}dz\int_0^{\infty}rdr \left\vert\phi_{0}\right\vert^2
    \end{equation}

  \begin{figure} [t]
    \centering
    \includegraphics[scale=0.26]{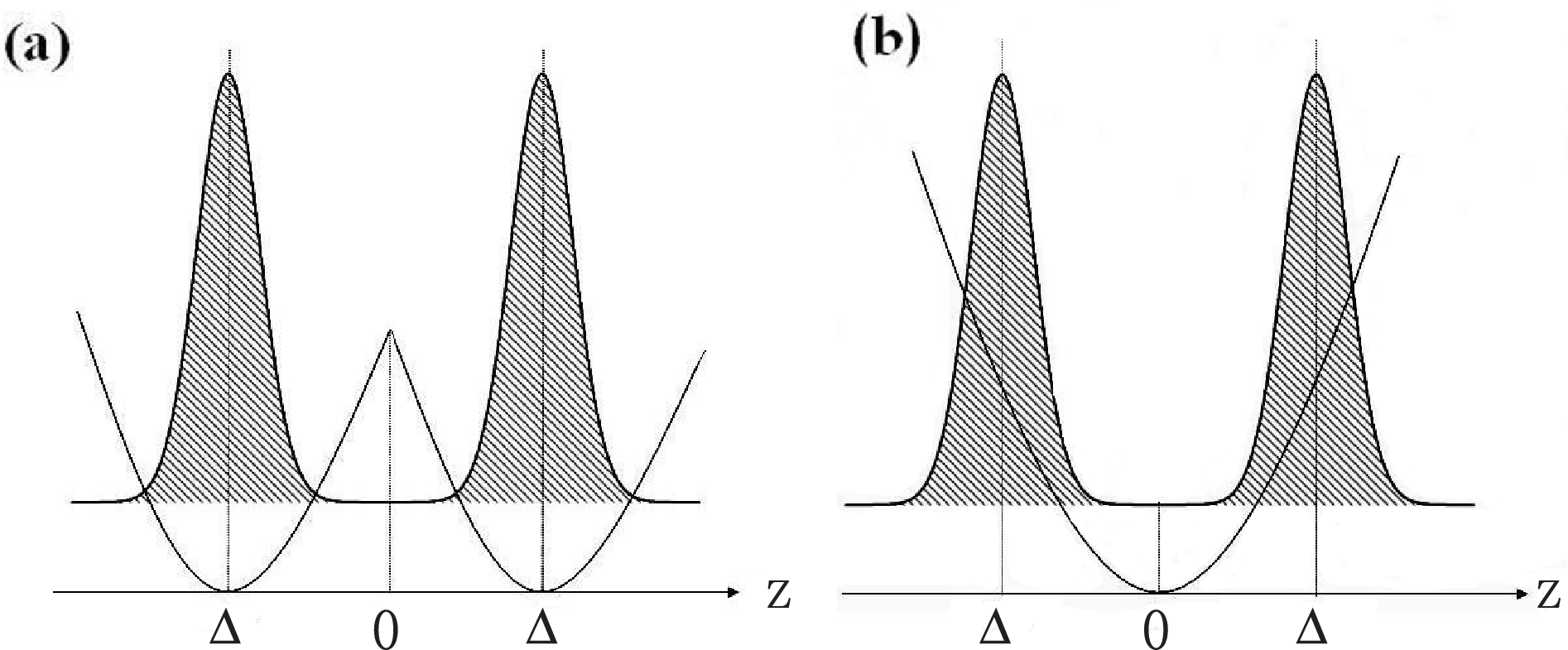}
    \caption{\label{Fig1} Solid curves: schematic representations of the effective potential for preparing two condensates (a) and for the interference process (b). Shaded areas in (a), (b) represent the density profile $|\phi(0,z;0)|^2$ of atom clouds.}
  \end{figure}

The initial state of a pair of clouds with displacement $\Delta$ in opposite directions along the $z$-axis, as shown in Fig.\ref{Fig1}(a), is then obtained by setting the order parameter to have the form
    \begin{equation} \label{Eq.5}
      \Phi({r},z;t=0) = \sqrt{\frac{1}{2\left(1+Q\right)}}\left(\phi_{0}({r},z-\Delta) + \phi_{0}({r},z+\Delta)\right).
    \end{equation}
This is a coherent superposition of two independent, normalized, condensate wave functions. The parameter $Q$ has the form
  \begin{equation} \label{Eq.6}
      Q=\frac{2\pi}{N}\int_{-\infty}^{\infty} dz \int_0^{\infty}rd{r}  \phi_{0} ( {r}, z-\Delta )  \phi_{0} ({r}, z+\Delta ),
  \end{equation}
which ensures correct normalization of $\Phi$ in situations in which there is a overlap between two atom clouds initially. It is worth pointing out that the representation in terms of a single order parameter, $\Phi$, implies that the two clouds are fully coherent and it is not the case that there are $N/2$ atoms in each cloud.

The subsequent evolution of the system in a new trap with frequencies $\omega_z$ and $\omega_\perp$, as shown in Fig.\ref{Fig1}(b), is studied by numerically solving the GPE
    \begin{equation} \label{Eq.7}
 i\hbar \frac{\partial\Phi}{\partial t} = -\frac{\hbar^2}{2m}\nabla^2 \Phi + V({r},z;t>0)\Phi + g\left\vert\Phi\right\vert^2\Phi\qquad,
    \end{equation}
assuming that the initial rotational symmetry about the $z$-axis is preserved. The dynamics of the system is fully determined by the trap potential used during the interference: expanding process with $\omega_{z}= \omega_{\bot} = 0$, colliding process with $\omega_{z} = \omega_{z}(0)$ and $\omega_{\bot}= \omega_{\bot}(0)$ and merging process with $\omega_{z}< \omega_{z}(0)$ and $\omega_{\bot}< \omega_{\bot}(0)$. To avoid the extra complexity may induced by the overlap of the clouds prepared in a double-well potential as described in Ref.\cite{PRA.87.023603} we will only discuss the situation where the two clouds are spatially well separated.

\section{Analytical approximation for merging condensates}

In absence of interatomic interaction there exists only kinetic energy and the interference fringes can be described by a linear Schr\"odinger equation. So the unperturbed condensates always show planar fringes during the interference.
If interatomic interaction $g\left\vert\Phi\right\vert^2\Phi$ is included the nonuniform density distribution is likely to lead to distinct rates of spatial expansion in the interfering condensates due to interaction-induced quantum pressure, thus resulting in position-dependent fringe spacings as well as curved fringes \cite{Andr}.

It is noted that if two condensates move along the $z$ direction, the interference fringe spacings are determined solely by the expansion rate $\eta_{z}$ and c.m. velocity $V_{z}$ in this direction, noted as $\lambda(\eta_{z}, v_{z}, t)$. In distinction to the c.m. velocity which has no contribution for the nonuniform and curved fringes, the position-dependent expansion rate of condensates arisen from the repulsive mean-field interaction is a critical factor for distorted fringes. Since for each slice of the trapped condensates along ${r}$, the fringe spacings $\lambda$ are associated only with $\eta_{z}$ and $v_{z}$, we can simplify the quasi-2D interference issues into 1D problems.

\begin{figure}[hbtp]
\centering
\includegraphics[scale=0.4]{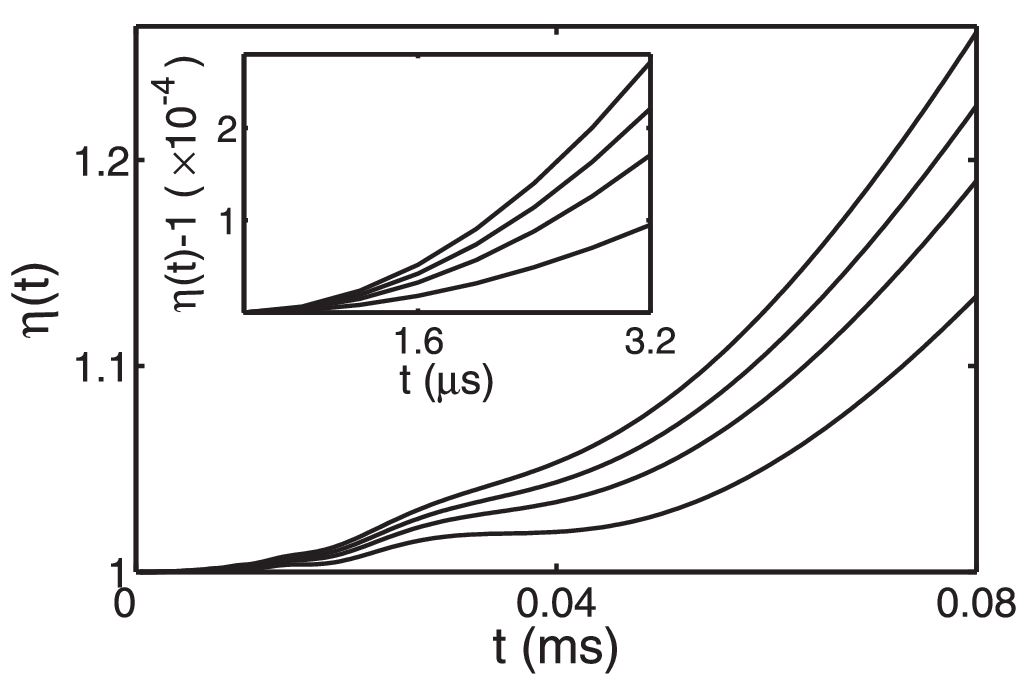}
\caption{\label{Fig2} The time-dependent expansion rate of the 1D freely expanding condensate, with $g_{1D}=1.3a_{s}$, from our simulation (curves from bottom to top have $N=$ 100, 400, 600, 800); Insert: short-time behaviors.}
\end{figure}

In the following, we attempt to find the generic relation between the mean-field interaction and the expansion rate of interference peaks of condensates by studying a 1D model whereby a single condensate initially confined in a harmonic trap experiences freely expansion. Conventionally the interference fringes of two condensates are with similar shapes but distinct peak densities. The nonuniform interactions induced by different peak densities cause the fringes to expand at different rates, and hence change its spacings. To match the features of interference fringes, we initially prepare different 1D condensates, depicted by $\Phi(z, t = 0)$, with different number of atoms but identical density profiles by adjusting harmonic trap frequency. Subsequently we drive these 1D condensates to expand freely by switching off the harmonic trap and explore their expansion rates, which are equivalent to those of the interference fringes of merging condensates. We quantify the rate of expansion by using $\eta=\int_{0}^{\infty}z|\Phi(z,t)|^{2}dz/\int_{0}^{\infty}z|\Phi(z,0)|^{2}dz$. Inspired by a 3D model of a classical gas \cite{cast}, we introduce a scaling factor $\eta (t)$ for our 1D case, where the condensate width $R(t)=\eta(t)R(0)$. The analytic solution of the expansion rate can be derived by adapting the method used in Ref.\cite{cast} to the 1D case. In the Thomas-Fermi approximation, $\eta(t)$ satisfies the dynamic equation \cite{cast}
  \begin{equation} \label{factor1}
    \ddot{\eta} =  \frac{\omega_z^{2}}{\eta^{2}},
  \end{equation}
where $\omega_z$ is the trap frequency. By integrating Eq.(\ref{factor1}), we find
  \begin{equation} \label{factor2}
    \frac{1}{2}\left( \sqrt{\eta(\eta-1)} +\textrm{log}\left|\sqrt{\eta}+\sqrt{\eta-1}\right| \right) =\frac{\omega_z t}{\sqrt{2}}.
  \end{equation}
For small $t$, the solution for Eq.(\ref{factor2}) is
  \begin{equation} \label{factor3}
    \eta \approx 1+\frac{1}{2}\omega_z^{2}t^{2}.
  \end{equation}
For large $t$, $\eta\gg 1$, and we obtain
  \begin{equation} \label{factor4}
    \eta \approx \sqrt{2}\omega_z t.
  \end{equation}
It is known that the peak density $n_p$ of 1D condensate increases with $\omega_z$. Eqs.(\ref{factor3}) and (\ref{factor4}) indicate that the expansion rate $\eta$ of 1D clouds also increases with $\omega_z$, which means that at a given time step the higher the peak density is the higher the expansion rate is. Fig.\ref{Fig2} shows that our simulations agree qualitatively with analytic prediction: the short-time behavior of $\eta(t)$ is approximately quadratic with $t$ while the long-time behavior is approximately linear with $t$.

The expansion rate of a Gaussian wavepacket in the absence of interactions can be described by $\frac{\hbar}{ml_{z}^{2}}$. By introducing an extra term $\frac{\alpha}{\hbar l_{z}^{2}}$, which is regarded as hydrodynamic pressure term in which $\alpha$ is assumed to be a parameter proportional to $gn_{p}$, we propose a phenomenological description of the effect of the repulsive interactions on the expansion of individual fringes.
Qualitative insight into the properties of the interference of two condensate can be gained by using a Gaussian ansatz,
	\begin{equation} \label{wave1}
	\phi(z;t) = C\sqrt{\frac{1}{1+i\chi(t)}}e^{-(1-i\chi(t))z^2/2\sigma^2(t)},
%\psi_{0}(x;t)=(2\pi l_{0}^{2})^{-1/4}\frac{l_{0}}{\sigma(t)}e^{-\frac{x^2}{4\sigma^2(t)}}
	\end{equation}
for the wave function, where $C=\left(\pi l_{z}^2\right)^{-1/4}$,  $\chi(t) = \hbar t/ml_{z}^2 +\alpha t/\hbar l_{z}^{2}$ and $\sigma^2(t) = l_{z}^2 \left(1+\chi^2(t)\right)$.
	
Now, we arrange a pair of well separated condensate wavepackets $\psi(z)$ initially at positions $\pm\Delta$ along the $z$-direction of the trap preparing them. The initial order parameter of the system is then
\begin{equation}
\Phi(z,t=0) =\frac{1}{\sqrt{2}} \left(\phi(z-\Delta,t=0)+\phi(z+\Delta,t=0)\right).
\end{equation}

If the clouds are released in another trap they will move toward each other with the same c.m. velocity $v=\hbar k/m$, with time-dependent order parameter
\begin{equation} \label{wave2}
   \Phi(z;t)=\frac{1}{\sqrt{2}}(\phi(z-\Delta;t)e^{-ikz}+\phi(z+\Delta;t)e^{ikz}).
\end{equation}
Inserting Eq.(\ref{wave1}) into Eq.(\ref{wave2}), the effective wavelength for the interference fringes can be deduced from the cosine term in $|\Phi(z;t)|^{2}$, which is
\begin{equation} \label{interference1}
  \lambda=\frac{2\pi[l_{z}^{4}+(\frac{\hbar t}{m}+\frac{\alpha t}{\hbar})^{2}]}{2\Delta(\frac{\hbar t}{m}+\frac{\alpha t}{\hbar}) +[l_{z}^{4}+(\frac{\hbar t}{m}+\frac{\alpha t}{\hbar})^{2}]2k'}.
\end{equation}
In absence of any expansion of the condensates (in the regime of colliding) as well as the interatomic interaction, the time-dependent terms in Eq.(\ref{interference1}) vanish and the associated fringe spacing reads,
\begin{equation}
\lambda=\frac{\pi}{k}=\frac{\pi\hbar}{mv}.
\end{equation}
When the clouds reach their maximum overlap the fringe spacing is $\lambda=\frac{h}{2m\omega_z(0)\Delta}$ analogous to classical interference.

If we suddenly turned off the trap the center of the clouds will not move ($k=0$), which means that the two wave packets expand freely and the interference fringe spacing then is
    \begin{equation} \label{Expanding}
      \lambda=\pi(l_{z}^{4}m+\frac{\hbar^2t^2}{m})/\hbar\Delta t
    \end{equation}
when $g_{1D}|\Phi|^2=0$. In general, the first term in the bracket in Eq.(\ref{Expanding}) can be neglected with respect to the second term at large $t$, resulting in the fringe spacing $\lambda= {ht}/{2m\Delta}$, which has been used in experiment \cite{Andr} to explain the interference of two expanding condensates.

Except for these limiting cases, all other processes of two condensates coming together are in the merging regime. Obviously the inter-atomic interaction affects the interference pattern in merging process. In this case the contribution of the free expansion term $\hbar t/m$ in Eq.(\ref{interference1}) can be ignored by comparing with that of the interaction term. Eq.(\ref{interference1}) then becomes
  	\begin{equation} \label{Collision}
      	\lambda\simeq\frac{\pi(l_{z}^{4}+\frac{\alpha^{2} t^{2}}{\hbar^{2}})}{\frac{2\alpha\Delta t}{\hbar} +(l_{z}^{4}+\frac{\alpha^{2} t^{2}}{\hbar})k},
  	\end{equation}
where $\lambda$ increases with increasing $\alpha$ ($\alpha\propto gn_p$).

\section{Distortion of fringes in two interfering condensates}

\begin{figure}[t]
        \centering
        \includegraphics[scale=0.48]{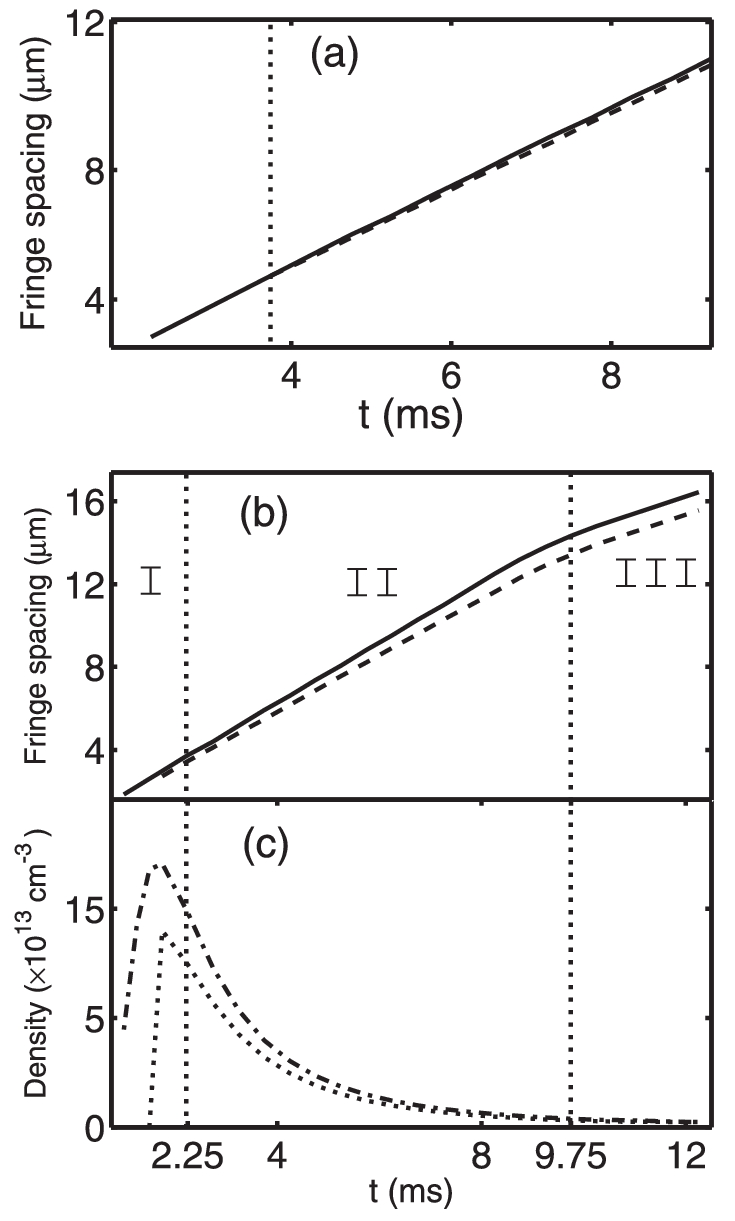}
        \caption{\label{Fig3} The zeroth-order (solid line) and the first-order (dashed line) fringe spacings of two freely expanding condensates versus time $t$ with $N=3\times 10^{4}$ (a) and $N=4\times 10^{5}$ (b). The time evolution of the peak densities of the zeroth- (dot-dashed curve) and first-order (dotted curve) fringes (c). Dotted line in (a) labels the time, $t=3.3$ms, at which the zeroth- and first-order fringes appear and dotted lines in (b) label the times that separate the three interference regimes. }
\end{figure}

In this section, we illustrate ``expanding'', ``colliding'' and ``merging'' behaviors of two BECs numerically. In particular, we focus on the role of the repulsive inter-atomic interactions in the distortion of fringes and the resulting spontaneous vortex formation.

\subsection{Free Expansion of Well Separated Clouds}

We set the initial separation of the two condensates to be $\Delta=4.8l_{z}$ with $l_{z}=\sqrt{\hbar/m\omega_{z}(0)}$ in a harmonic trap with frequencies $\omega_{z}(0)= 2\pi\times 180 \text{Hz}$ and $\omega_{\perp}(0) = 2\pi\times 120\text{Hz}$. Then, the system is allowed to evolve freely by setting $\omega_{z}=\omega_{\perp}=0$. In the absence of interactions, the order parameter of a single wave packet can be described by \cite{pita}
      \begin{equation} \label{Eq.13}
         \begin{split}
           \phi({r},z;t) &= \left(\frac{m\overline\omega}{\pi\hbar}\right)^{3/4} \alpha_{\perp}(t) \sqrt{\alpha_{z}(t)} \\
                       & \times e^{-\left(\alpha_{z}(t)-i\beta_{z}(t)\right)z^2/2l_{z}^2}                                                                       e^{-\left(\alpha_{\perp}-i\beta_{\perp}\right)r^2/2l_{\perp}^2}
        \end{split}
      \end{equation}
with $\alpha_j  =  {1}/\left({1+\omega_j^2 t^2}\right)$ and $\beta_j  =  {\omega_j t}/\left({1+\omega_j^2 t^2}\right)$.
According to Eq.\ref({wave2}), at arbitrary time $t$ we have
      \begin{eqnarray}
        n({r},z;t) & = & \left\vert\Phi({r},z;t)\right\vert^2 \nonumber \\
                 & = & \frac{2\alpha_{\perp}^2(t) \alpha_{z}(t)}{\pi^{3/2} l_{\perp}^2 l_{z}} e^{-\alpha_{\perp}(t)r^2/l^2_{\perp}}                            e^{-\alpha_{z}(z^2+\Delta^2)/l^2_{\perp}} \nonumber \\
                 &\times & \left(\cosh\left(\frac{2\alpha_{z}(t)\Delta                                       z}{l_{z}^2}\right)+\cos\left(\frac{2\beta_{z}(t)\Delta z}{l_{z}^2}\right)\right) \label{fringe1}\qquad
      \end{eqnarray}
From the cosine term in Eq.(\ref{fringe1}), we deduce that the length scale characterizing the interference fringes is
\begin{equation} \label{Eq.16}
 \lambda(t) = \frac{\pi l_{z}^2}{\Delta}\frac{1+\omega^2_{z}(0) t^2}{\omega_{z}(0) t}
\end{equation}
which, for $t\gg 1/\omega_{z}(0)$, behaves as
\begin{equation}
\lambda(t)\sim (\pi l^2_{z}/\Delta)\omega_{z}(0) t=\frac{h}{2m\Delta} t,
\end{equation}
which is the same as we derived from Eq.(\ref{Expanding}) of the 1D model.

\begin{figure}[t]
        \centering
        \includegraphics[scale=0.6]{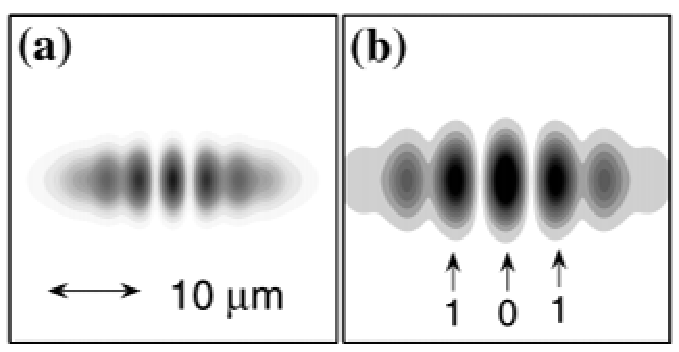}
        \caption{\label{Fig4} Gray-scale plots of atom density (black=high) in the $z$-$r$ plane (axes inset) for interfering condensates, evolving with the process of expanding at $t=5.81$ms(a) and $10.21$ms(b). $0$ and $1$ label zeroth- and first-order fringes, respectively.}
\end{figure}

When the peak density of the system is low the nonlinear effect is not obvious. As shown in Fig.\ref{Fig3}(a), at a relatively low density ($N=3 \times 10^{4}$), the growth of the fringe spacing for the zeroth- and first-order fringes (depicted in Fig.\ref{Fig4}) is linear in time, which is similar to the the non-interacting expression we discussed above. The growth rates are nearly the same for short-time evolution ($t<3.3ms$), while there is only slightly difference for $t>3.3ms$. At a higher density ($N= 4\times 10^{5}$), the interference dynamics show distinct properties in three regimes, as indicated in Fig.\ref{Fig3} (b). In regime I ($t<2.25\text{ms}$), the growth of the fringe spacings is identical for the zeroth- and first-order fringes due to the low density at the beginning of the interference. Note that this is different from the non-interacting behavior because the nonlinear interactions increase the expansion rate of the clouds \cite{wall}. In regime II ($2.25$ms$<t<9.75\text{ms}$), the spacing of the zeroth-order fringe grows faster than that of the first-order fringes, indicating that the ideal fringe pattern is distorted in the $z$-direction. Specifically, owing to its higher density, the central fringe is thickened relative to its neighbors. In regime III ($t>9.75$ms), the growth rate of the zeroth- and first-order fringe spacings reduces gradually, approaching the non-interacting limit. For enough long times, the difference in the zeroth- and first-order fringe spacings becomes negligible and the fringes are uniform.

\begin{figure}[t]
        \centering
        \includegraphics[scale=0.5]{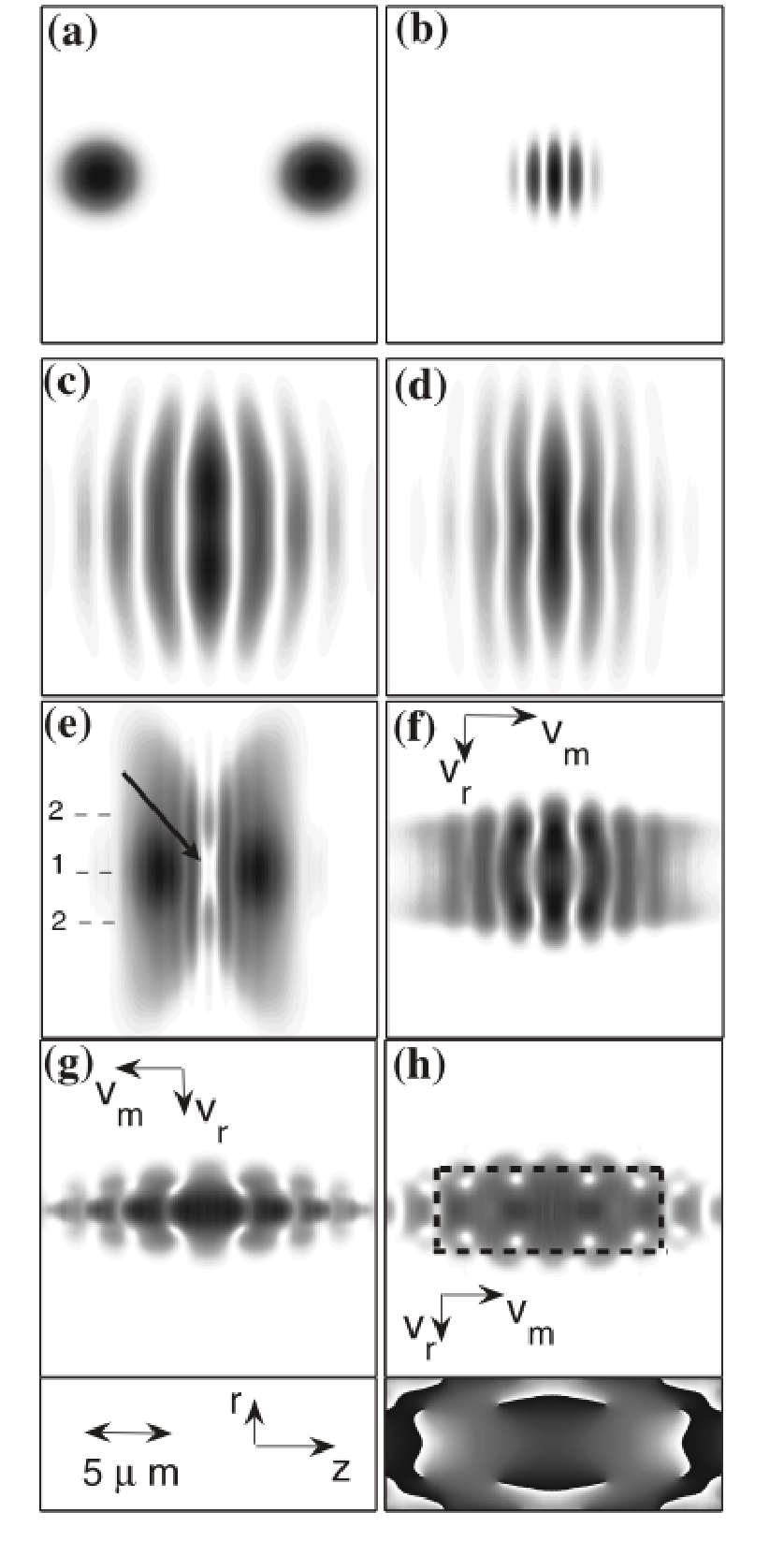}
        \caption{\label{Fig5} Gray-scale plots of atom density (black=high) in the $z$-$r$ plane (axes inset) for
        double condensates, evolving with the process of colliding at $t=0$ms (a), $2.01$ms (b), and merging at
        $t=1.21$ms (c), $2.01$ms (d), $2.41$ms (e), $3.61$ms (f), $4.11$ms (g), $4.41$ms (h); The phase of the BEC
        wave function within the dashed box in (h) is shown at the bottom of the density profile (black=$2\pi$, white=$0$).
        }
        %arrows $v_{m}$ and $v_{r}$ show the directions of motion in left-upper(bottom) part of clouds.}
\end{figure}

\subsection{Collision of Well Separated Clouds}

In this set of simulations, the initial state is prepared as before for a range of value of $N$. However, the trap remains present during the dynamical evolution, as depicted in Fig.\ref{Fig1}(b), so that the two clouds are accelerated towards one another. For low-density condensates at a large displacement, the system behaves, at least over one period of the longitudinal trap, like an ideal Bose gas. Since we keep the trap frequencies unchanged, the initial state is a superposition of two coherent states of the oscillator potential, so that
\begin{equation} \label{Eq.10}
        \begin{split}
          \Phi({r},z;t) & = \frac{1}{\sqrt{2}}e^{-{r}^2/2l_{\perp}^2}\left(e^{-iK(t)z} e^{-(z-Z(t))^2/2l_{z}^2} \right. \\
          & \left.+ e^{iK(t)z} e^{-(z+Z(t))^2/2l_{z}^2}\right)\qquad
        \end{split}
\end{equation}
where $Z(t)=\Delta \cos \left(\omega_{z} t\right)$ and $K(t) = {\Delta}\sin\left(\omega_{z} t\right)/{l_{z}^2}$.

\begin{figure}[t]
\centering
\includegraphics[scale=0.5]{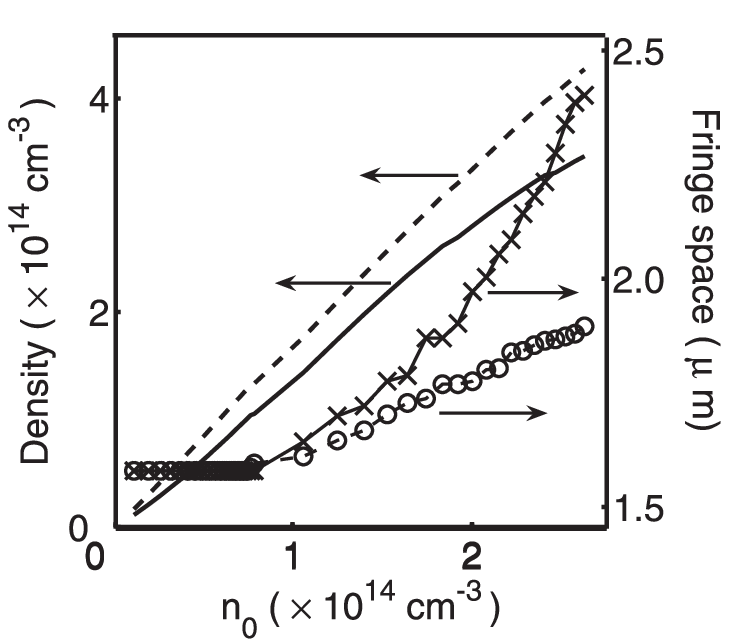}
\caption{\label{Fig6}The peak density of the zeroth- (dashed curve) and first-order fringes (solid curve),  and the zeroth- (solid curve with crosses) and first-order fringe width (dashed curve with circles) versus the initial peak density of the condensate, $n_{0}$, in ``colliding'' process with fixed $g$.}
\end{figure}

The initial density distribution of the system is shown in Fig.\ref{Fig5}(a). Subsequently the two clouds are released in the trap and approach each other. The clouds are maximally overlapping at time $\bar{t}=\pi/2\omega_{z}$ when they reached the bottom of the trap. The corresponding c.m. velocity $\bar v$ and the fringe spacing $\lambda$ are given by
\begin{eqnarray} \label{Eq.11}
        \overline{v}&=&\omega_{z}\Delta \qquad ,\\
        \bar\lambda &=& \frac{2\pi}{K\left(\bar{t}\right)} =\frac{2\pi\hbar}{m\bar v}=\frac{2\pi l_{z}^2}{\Delta}\qquad.
\end{eqnarray}
The collision velocity and the fringe spacing under the given parameters are $\bar{v}=6.08 l_{z}\omega_{z}$ and $\bar\lambda =1.036 l_{z}$, respectively. In Fig.\ref{Fig5}(b) we can see that interference fringes are nearly straight within the first half period of c.m. oscillation. It is due to the fact that at a relative high peak c.m. velocity $v$ the interaction induced scattering is not effective due to the short interference time. With increasing number of atoms (increasing $n_0$) the interactions are more important during the interference. Fig.\ref{Fig6} shows the effect of the nonlinear term on the fringe pattern of the colliding condensates with different atom numbers. For an initial peak density $n_{0}<1.52 \times 10^{14}$cm$^{-3}$, the peak density of the zeroth- and first-order fringes increases linearly with atom number, as expected for interference in a system governed by a linear wave equation. In this regime, for given initial trap frequencies and separation of the clouds, the fringe spacings are \emph{independent} of the initial density distribution. However, when $n_0>1.52 \times 10^{14}$cm$^{-3}$, the peak densities of the first two fringes depend sublinearly on $n_{0}$. The gap between the two fringes increases with $n_{0}$, with the result that the differences between fringe spacings increase as well. When this difference in the high-density region becomes sufficiently large, the fringes become significantly curved leading, ultimately to vortex formation. In Ref.\cite{PRA.87.023603} we showed that with the same number of atoms but smaller displacement, especially when the peak c.m. kinetic energy is the same order as the interaction energy ($\Delta\sim R_{TF}$), the distortion of the fringes become quite sizable and after the first interference the two clouds can not separate clearly. The next interfering is then dominated by the interatomic interactions with soliton and vortex excitations.

Furthermore, our phenomenological formula Eq.(\ref{Collision}) gives a good description of the nonuniform fringes. The higher density in the zeroth-order fringe, characterized by larger $\alpha$, leads to a larger fringe spacing than that of the first-order fringe.

\subsection{Merging of Well Separated Clouds}

To study the merging process, the condensates are originally prepared in a harmonic trap with frequencies $\omega_{z}(0) = 2\pi \times 800$Hz and $\omega_{\perp}(0) = 2\pi \times 533$Hz. The total number of atoms is $N=3\times 10^4$ with peak density of the ground state clouds being $n_0=4.11\times 10^{15}\text{cm}^{-3}$. We chose the c.m. displacement, $\Delta = 2.8l_z$, which is smaller than that used above to strengthen the role of interactions.  At time $t=0$, we suddenly change the trap frequencies into $\omega_{z} = 2\pi\times 180 \text{Hz}$ and $\omega_{\perp} = 2\pi\times 120\text{Hz}$. Since these frequencies are much smaller than those used for preparing the clouds, each cloud expands due to internal pressure as well as undergoing bulk oscillation, which allows us to investigate the effect of the combination of expansion and oscillation of clouds on the dynamics of the system.

Fig.\ref{Fig5}(c)-(h) show the time evolution of the density profile. The process of both expansion and collision leads to a distorted fringe pattern with the characteristic configuration that there are ``thicker'' fringes in the center of the cloud and a larger fringe spacing, towards the edge of the cloud where interaction effects are negligible and the fringe spacing is closer to its non-interacting value (Fig.\ref{Fig5}(c)). This curved fringe pattern is similar to that observed in experiment \cite{Andr} where condensates with small separation tend to form large variations in density in the merging region. After the clouds have completely merged, the fringe spacings become smaller, and the distinction between the center and edge part as well as the zeroth-order and first-order peak, reduces correspondingly (Fig.\ref{Fig5}(d)). The most significant phenomenon is that along $r=0$ the zeroth-order fringe spacing is not the widest part as seen in Fig.\ref{Fig5}(c) and \ref{Fig5}(d). This is quite different from the zeroth-order fringe formed in the critically damped regime where there is only bulk c.m oscillation \cite{PRA.87.023603} which takes a lenticular shape. This kind of distortion is induced by the expansion of the clouds, which makes the scattering between two clouds in the $r$-direction sufficient. From Figs.\ref{Fig5}(c)-(e) we can see the expansion of the clouds in the radial direction is sizable and the self-interference in the radial direction in the following collisions will increase the instability of the system. This eventually causes decay of interference fringes via snake instability \cite{PRL.76.2262} leading to the spontaneous vortex formation as shown in Fig.\ref{Fig5}(h).

\begin{figure}[hbtp]
\centering
\includegraphics[scale=0.58]{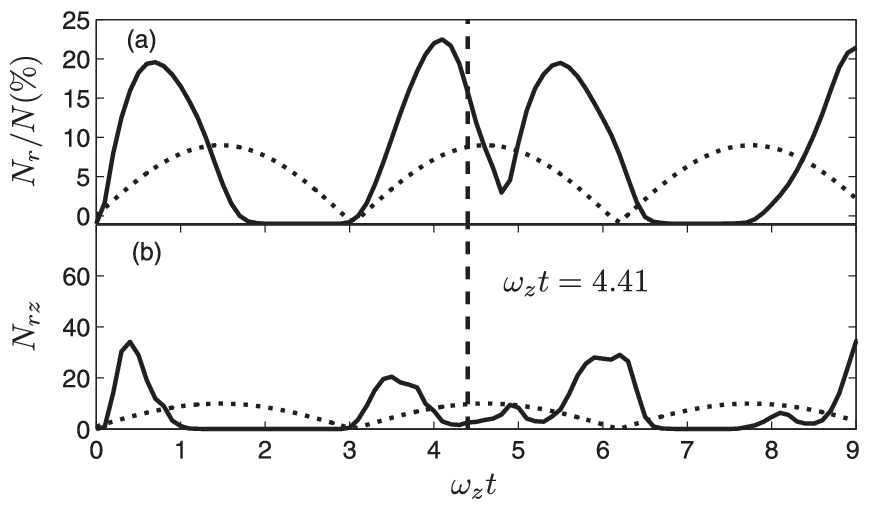}%[width=8.5cm, height=7cm]{Fig7.eps}
\caption{\label{Fig7}Oscillation of condensate atom number $N_r$ (a) and $N_{rz}$ (b). The dotted line is $\sin(\omega_z t)$ with arbitrary amplitude.}
\end{figure}

The complex interference pattern can also be interpreted by Eq.(\ref{interference1}). At the beginning of the merging ($t<2.1$ms), the expanding speed of an atom cloud $v_e$ is larger than its c.m. speed, $v_m$, due to the very large $n_{0}$, and thus the fringe spacings are dominated by the term $\alpha t/\hbar$ in Eq.(\ref{interference1}). This implies that the larger peak density of the fringes, the larger $\alpha$, resulting in larger $\lambda$. Therefore, the fringe spacings approaching $z=0$, are slightly larger than those at the edge, because the peak density of the zeroth-order fringe is larger than those of higher-order fringes. With time evolution in the merging process, the increasing $k$ and $\hbar t/2m$ in Eq.(\ref{interference1}) gradually dominate, resulting in smaller fringe spacings and less variation of the fringe spacings between the center and edge of the clouds.

As shown in Fig.\ref{Fig5}(d) and (e), the strong interactions in the center of the clouds, lead to a net radial
flow of atoms in the high density fringes, triggering a sound wave in the central peak which propagates radially
from position $1$ to $2$. Our calculation shows that $v_{r}\approx L_{1,2}/t_{1,2}=11.1$mms$^{-1}$, larger than
$2v_{m}$ ($\approx9.8$mms$^{-1}$). This rapid outflow leads to complete depletion of the center of trap after the two clouds have passed through one another, resulting in the formation of a dark soliton (see Fig.\ref{Fig5}(e)). When the two clouds recollide at time $2\overline{t}$, both are recollapsing radially leading to an enhanced density in the bright fringes causing an even greater longitudinal expansion and greater density gradients in the radial direction (see Fig.\ref{Fig5}(f)). The curvature of the fringes is thus even greater than in the first collision, as shown in Fig.\ref{Fig5}(g), and is sufficient to generate a net circulation around localized regions that are fully depleted: vortices are generated, as seen in Fig.\ref{Fig5}(h).

In Fig.\ref{Fig7}, we show quantitatively the scattering effect by plotting the dynamics of $N_r$, which is the number of atoms with $r$-component of momentum $k_r=\sqrt{k_x^2+k_y^2}>20dk_r$ where $2\pi/20.05l_z$ is the spatial resolution of the simulation. $N_{rz}$ is the number of atoms with $k_r=\sqrt{k_x^2+k_y^2}>20dk_r$ and $k_z>40dk_z$. In Fig.\ref{Fig7}(a) the first and the third peak are induced by the interference of the clouds in the first period of the c.m. oscillation, while the second one arises from the self-interference of the clouds in the $r-$direction. The interference starts very soon after the clouds are released in the trap and the value of $N_r$ increases up to $20.5\%$ rapidly, and the second peak is higher than the third one, which highlights the fact that the expansion enhances the scattering effect. In absence of expansion (Ref.\cite{PRA.87.023603}) the scattering effect is weaker than here so that the second peak is much lower than the third one. As shown in Fig.\ref{Fig7}, at $t=4.41/\omega_z$ the spontaneous vortex generation (see Fig.\ref{Fig5}(h)) highly depresses the number modulation.

In general, due to strong inter-atomic interaction and the competition between $v_{m}$ and $v_{r}$, additional degrees of freedom are generated, with the result that the high-density areas from the interference peaks encircle the low-density areas between them (Fig.\ref{Fig5}(g)). The positions of the vortex rings correlate naturally with the previous positions of interference valleys. Naturally, the formation of vortices from quantum reflection of high-density and low-velocity BEC \cite{Scott} is analogous with our results.

\section{General Features in the parameter space}
\begin{figure}[hbtp]
\centering
\includegraphics[scale=0.35]{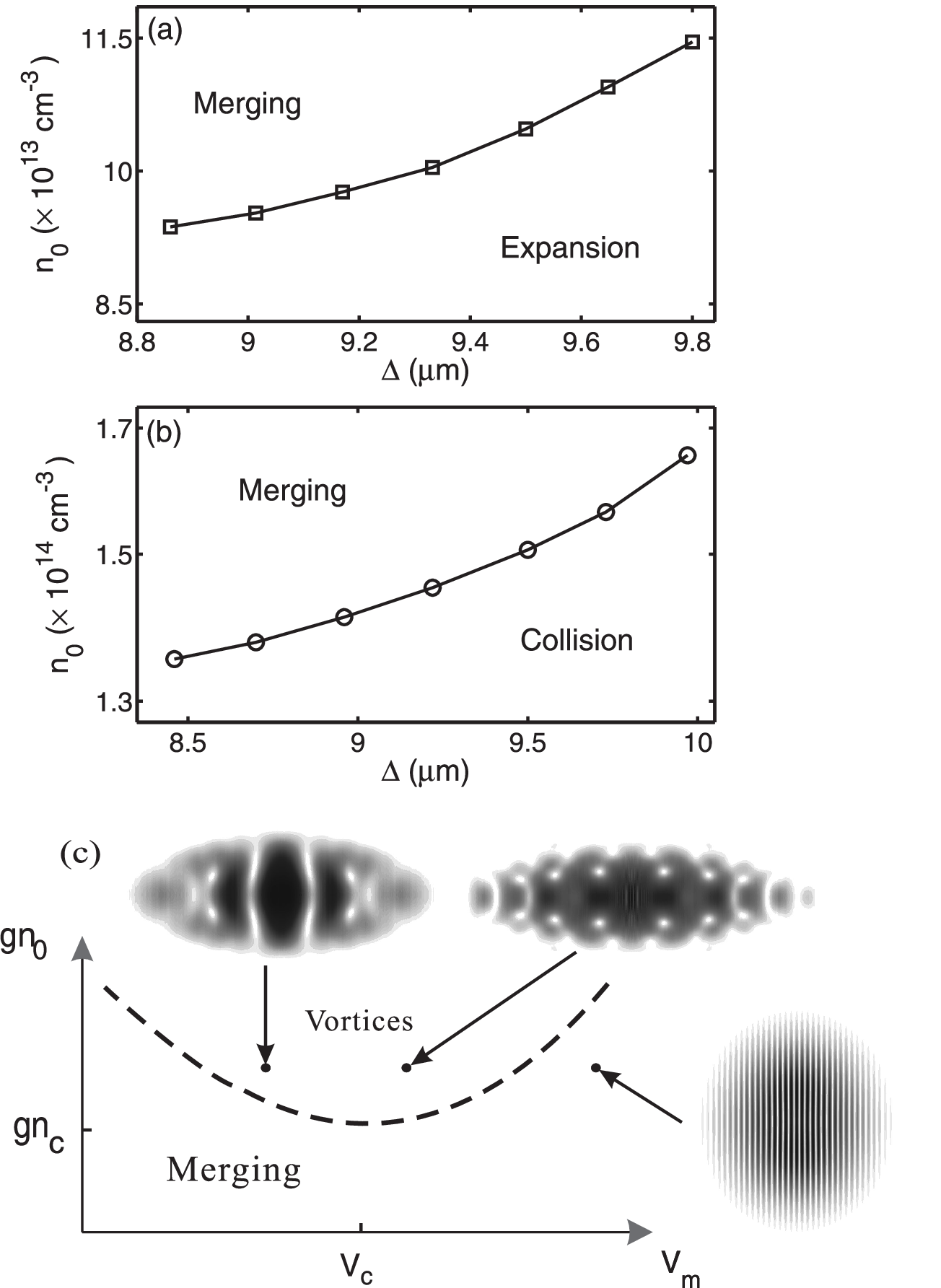}
\caption{\label{Fig8} In (a) and (b) the cross-overs between different qualitative behaviors are illustrated. The generic ``merging'' process is identified by a fringe non-uniformity of at least 0.4$l_{z}$ between zeroth- and first-order fringes in case (a) and at least 10\% in case (b). In (a) a pair of identical clouds are placed a distance $\Delta$ apart and allowed to expand in the absence of any trap potential. In (b) the clouds evolve in a single trap potential. In (c) the locus of points in the parameter space above which vortices are produced is shown schematically for a generic process. Insert: three density configurations for the same clouds with different separations in the merging and overlapping process with their parameter points arrowed.}
\end{figure}

To explore the interference properties and characterize the spontaneous formation of vortices over the parameter space, we have performed a comprehensive set of numerical simulations to obtain Fig.\ref{Fig8}. Each data point in Fig.\ref{Fig8} (a) and (b) is derived from a set of at least 12 dynamic calculations. We identify the three types of process through their interference characteristics. In an expansion process, the interference fringes are time-dependent and uniform. In the colliding process, the fringes are static and uniform while the clouds are fully overlapping. The interference pattern in a merging process is time-dependent and nonuniform.

Fig.\ref{Fig8} (a) demonstrates that starting in the freely expanding regime, moving to higher initial densities or smaller initial separations will produce the more general merging behavior. This is exhibited by a loss of homogeneity in the fringe spacing and, in turn, fringe curvature. The upwards curve of the boundary between expanding and merging behavior is similar to that for the boundary between colliding and merging behavior shown in (Fig.\ref{Fig8} (b)).

Based on the analysis of the dynamics over a large portion of the parameter space, Fig.\ref{Fig8} (c) summarizes the formation of vortices in merging BECs. Above a critical initial density, $n_c$, vortices will form during the second overlap of the two clouds within a definite range of initial c.m. velocities. If the velocity is too low then damping of the motion during the first overlap effectively arrests the dynamics. The low velocity leads to a large fringe spacing so that, even towards the edges of the cloud, the curvature is never sufficient to cause a net circulation. 

If the collision velocity is too high, the hydrodynamic response of the cloud is too slow and the clouds have separated before any significant distortion occurs \cite{scott2} (inserted plot in Fig.\ref{Fig8} (c)). Our simulations show that the curvature of the interference fringes is of importance in the formation of vortices and that the number of fringes within the cloud determine the number of vortices. This is shown in the inserted plots of Fig.\ref{Fig8} (c) where the five fringes from the short initial separation result in the formation of two vortex rings while the nine fringes from the large initial separation cause six vortex rings.
The connection between the distortion of fringes and the formation of vortices can explain the experiments in which more vortices are generated in the faster merging, interfering region \cite{davi}. The faster merging produces more interference fringes, creating the possibility for the formation of more vortices in the region. The diagram in Fig.\ref{Fig8} is based on the specific material parameters described in the previous sections. Simulations using different parameters give results that are similar but with the cross-over lines in Fig.\ref{Fig8} shifted. In general, for the smaller values of the coupling constant, $g$, and with the larger numbers of atoms, it is easier to observe the crossover from colliding process to merging process.

Finally, we address the effect of the halo of scattered atoms, produced by the counter-propagating condensates, on the interference pattern. The phenomenon that if two BECs collide at a sufficiently high velocity, a halo of elastically scattered atoms is produced, has been demonstrated by both experiments \cite{chik, voge, katz} and theory \cite{norr, chwe}. In the expanding and merging processes of our system, the scattered atoms might not have significant effect on the interference patterns because the rapid decrease of the densities corresponding to the fast expansion of the clouds would reduce the rate of atom-atom collision in merging areas. For the colliding process, the visibility of the interference can be reduced and the fringes can be distorted due to the scattered atoms. However, these effects can be suppressed or avoided if the density and velocity of clouds are low, or the colliding process is controlled by properly adjusting the barriers \cite{davi, Jo, Jo1}.

\section{Summary}
In summary, we have investigated the distortion of interference fringes and vortex formation in two
merging condensates and identified three distinct regimes in situations relevant to atom interferometry.
Our simulations provide an explanation of recent experimental work \cite{davi}, where the faster mergence of three BECs creates more vortices.
The regularity of vortex formation and interference might allow for the design of experiments to study vortex creation and the dynamics of regular, linear, vortex arrays. Our latest calculations show clearly that one could control the density of vortices generated by tuning the fringe spacing.

\section*{ACKNOWLEDGMENTS}
We thank R. G. Scott and T. M. Fromhold for fruitful discussion. We also acknowledge support by the EPSRC.

\end{document}